\documentstyle[aps,prl,preprint,floats,epsfig]{revtex}

\def\pio{\pi^{0}}
\def\pgamma{\tau^{-} \to \bar{p} \gamma}
\def\ppio{\tau^{-} \to \bar{p} \pio}
\def\peta{\tau^{-} \to \bar{p} \eta}
\def\p2pio{\tau^{-} \to \bar{p} 2 \pio}
\def\ppioeta{\tau^{-} \to \bar{p} \pio \eta}

\def\BR{{\cal B}}
\newcommand{\zpc}[3]{Z. Phys. C {\bf #1}, #2 (#3)}
\def\etal{{\em et~al.}}

\def\effa{ 10.7 \pm 0.2 }
\def\effb{ 8.4 \pm 0.2 }
\def\effc{ 14.0 \pm 0.2 }
\def\effd{ 4.3 \pm 0.1 }
\def\effe{ 4.6 \pm 0.1 }

\def\oba{1}
\def\obb{14}
\def\obc{2}
\def\obd{41}
\def\obe{1}

\def\bga{6.0}
\def\bgb{13.5}
\def\bgc{4.0}
\def\bgd{50.5}
\def\bge{0.5}

\def\bra{3.5}
\def\brb{15}
\def\brc{8.9}
\def\brd{33}
\def\bre{27}

\begin{document}

\preprint{\tighten\vbox{\hbox{\hfil CLNS 98/1595}
                        \hbox{\hfil CLEO 98-19}
}}

\title{Search for Baryon and Lepton Number\\
       Violating Decays of the $\tau$ Lepton}  

% Your author list ***DOES NOT*** go here!
% is goes below where you are instructed to insert it...
 
\author{CLEO Collaboration}
\date{\today}

\maketitle
\tighten

\begin{abstract} 
We have searched for five decay modes of the $\tau$ lepton that simultaneously
violate lepton and baryon number: $\pgamma$, $\ppio$, $\peta$, 
$\p2pio$, and $\ppioeta$. 
The data used in the search were collected with the
CLEO~II detector at the Cornell Electron Storage Ring (CESR).
The integrated luminosity of the data sample is 4.7 fb$^{-1}$,
corresponding to the production of $4.3 \times 10^6\ \tau^+\tau^-$ events.
No evidence is found for any of the 
decays, resulting in much improved upper limits on the branching
fractions for the two-body decays and first upper limits for
the three-body decays.
\end{abstract}
\newpage

{
\renewcommand{\thefootnote}{\fnsymbol{footnote}}

\begin{center}
R.~Godang,$^{1}$ K.~Kinoshita,$^{1,}$%
\footnote{Permanent address: University of Cincinnati, Cincinnati OH 45221}
I.~C.~Lai,$^{1}$ P.~Pomianowski,$^{1}$ S.~Schrenk,$^{1}$
G.~Bonvicini,$^{2}$ D.~Cinabro,$^{2}$ R.~Greene,$^{2}$
L.~P.~Perera,$^{2}$ G.~J.~Zhou,$^{2}$
% ...
\end{center}
 
\small
\begin{center}
$^{1}${Virginia Polytechnic Institute and State University,
Blacksburg, Virginia 24061}\\
$^{2}${Wayne State University, Detroit, Michigan 48202}\\
% ...
\end{center}

\setcounter{footnote}{0}
}
\newpage

Baryon and lepton number conservation are experimentally observed phenomena.
In the Standard Model, both numbers are assumed to be conserved.
Baryon and lepton number violations are expected in many extensions of
the Standard Model such as supersymmetry and superstring
inspired models~\cite{BL}.
In some of the models, there is a new symmetry associated with the 
conservation of the baryon minus lepton number, 
$B-L$, even though both baryon and lepton numbers are not conserved.
Decays with this new symmetry have been searched for in
nucleon decays~\cite{pdg}.
The lower limits on the nucleon lifetimes imply that
the corresponding decays involving the $\tau$ lepton
are below the current experimental sensitivity~\cite{Marciano}.
Nevertheless, experimenters have searched for this type of decay
because the $\tau$ lepton provides a clean laboratory for the search~\cite{proton}.
The previous upper limits~\cite{ARGUS} on the branching fractions
for the decays into an anti-proton~\cite{charge} and 
a photon, $\pi^0$ or an $\eta$ 
meson are in the range of $10^{-4}-10^{-3}$.
There are no published results for the decays involving two neutral mesons, 
$\p2pio$ and $\ppioeta$.
The CLEO~II experiment, with its large sample of $\tau$ events,
provides an opportunity to search for decays that violate
lepton and baryon numbers, but conserve $B-L$.
In this paper, we present the result of a search in
five decay modes: $\pgamma$, $\ppio$,
$\peta$, $\p2pio$, and $\ppioeta$. 

The data used in this analysis were collected with the CLEO~II
detector from $e^+e^-$ collisions at the Cornell Electron
Storage Ring (CESR) at a center-of-mass energy $\sqrt{s} \sim 10.6$ GeV.
The total integrated luminosity of the data sample is 4.7 fb$^{-1}$,
corresponding to the production of
$N_{\tau\tau} = 4.3 \times 10^6\ \tau^+\tau^-$ events.
CLEO~II is a general purpose spectrometer~\cite{cleoiinim} with
excellent charged particle and shower energy detection.
The momenta of charged particles are measured with three
drift chambers between 5 and 90~cm from the
$e^+e^-$ interaction point (IP), with a total of 67 layers.
The specific ionization $(dE/dx)$ of charged particles is also
measured in the main drift chamber.
These are surrounded by a scintillation time-of-flight
system and a CsI(Tl) calorimeter with 7800 crystals.
These detector systems are installed inside a
superconducting solenoidal magnet (1.5~T), surrounded by
an iron return yoke instrumented with proportional tube chambers
for muon identification.
 
The $\tau^+\tau^-$ candidate events 
must contain two charged tracks with zero net charge. 
To reject beam-gas events, the distance of closest approach
of each track to the IP must be within 0.5 cm transverse
to the beam and 5 cm along the beam direction.
Photons are defined as energy clusters in the calorimeter
with at least 60 MeV in the barrel ($|\cos\theta| < 0.80$)
or 100 MeV in the endcap ($0.80 < |\cos\theta| < 0.95$),
where $\theta$ is the polar angle defined with respect
to the beam axis.
We further require every photon to
be separated from the projection of any charged track on
the surface of the calorimeter by at least 30 cm 
unless its energy is greater than 300 MeV. 

We divide each event into two (signal and tag) hemispheres, 
each containing one charged track,
using the plane perpendicular to the thrust axis~\cite{thrust},
which is calculated from both charged tracks and photons.
The charged track in the tag (signal) hemispheres is assumed to be a
pion (an anti-proton).
The invariant mass of the particles in the tag hemisphere
must be less than the $\tau$ mass, 
$M_{\tau}=1.777$~GeV/c$^2$~\cite{pdg}.
To suppress the background from radiative Bhabha and $\mu$-pair events,
the direction of the missing momentum of the event
is required to satisfy $|\cos\theta_{missing}| < 0.90$,
where $\theta_{missing}$ is the angle of the missing momentum
defined with respect to the beam axis. 
Because there is no neutrino in the signal hemisphere
while there is at least one neutrino undetected in the tag hemisphere,
the missing momentum of the event must point toward the tag hemisphere,
$0 < \cos \alpha < 1.0$, where $\alpha$ is the angle between the missing
momentum and the total momentum of the particles in the tag hemisphere.

Several additional selection criteria are imposed on the
decays $\pgamma$ and $\ppio$ to suppress the background.
For the decay $\pgamma$, we further impose
the restriction $\cos \alpha < 0.99$
to reduce the background from radiative Bhabha and $\mu$-pair events.  
The background is further reduced by requiring the net transverse momentum
of each event with respect to the beam axis to be greater than 300 MeV/c.
The Bhabha background is further suppressed by rejecting
events with an electron in the tag
hemisphere. An electron is defined as a particle having a
shower energy-to-momentum ratio with $E/p > 0.85$ and a
specific ionization loss $(dE/dx)$ within 3 standard deviations
of the expectation.
The migration background from other $\tau$ decays is
suppressed by restricting the angle
between the momentum vectors of the $\bar{p}$ and $\gamma$,
$0.35 < \cos\theta_{\bar{p}\gamma} < 0.92$.
For the decay $\ppio$, the $\bar{p}$ 
momentum must be greater than 2.5 GeV/c to reduce further the 
background from $\tau$ migration. 

We reconstruct $\pi^0$ and $\eta$ mesons with photons in
the barrel using the 
$\gamma \gamma$ decay channel. In order to maintain a high
detection efficiency while minimizing the dependence
on the Monte Carlo simulation of electromagnetic showers, 
there is no explicit cut on the maximum number of photons
in the signal hemisphere. However, photons that are most
likely to be real must be used in the signal decay reconstruction.
These are photons passing the 30~cm isolation cut and
having either an energy above 300~MeV or a lateral profile of
energy deposition consistent with that expected of a photon. 

The $\gamma \gamma$ invariant mass spectrum is expressed in standard deviations
from the nominal $\pi^0$ or $\eta$ mass,
\begin{eqnarray*}
S_{\gamma \gamma} = (M_{\gamma \gamma}-M_{\pi^0,\eta})/\sigma_{\gamma \gamma}\ \ ,
\end{eqnarray*}
\noindent where $\sigma_{\gamma \gamma}$ is the mass resolution calculated
from the energy and angular resolution of each photon. 
The signal region is defined as $-3 < S_{\gamma \gamma} < 2$;
the asymmetric cut is used to account for shower leakage.

To search for the decay candidates, we select
$\tau$ candidates with invariant mass and total energy
consistent with expectations.  The following kinematic 
variables are used to select the candidate events:
\begin{eqnarray*}
\Delta E & = & E - E_{beam} \\
\Delta M & = & M - M_{\tau}\ \ ,
\end{eqnarray*}
\noindent where $E_{beam}$ is the beam energy, and $E$ and $M$ 
are the reconstructed $\tau$ candidate energy and mass, respectively.  
The decay candidates are required 
to have both kinematic variables within $1.28\sigma$  of the expectations
(80\% efficiency for each variable). 
For the decays involving $\eta$
mesons, which have a smaller $\tau$ migration background, 
the requirement is loosened to $1.64\sigma$ (90\%~efficiency for 
each variable).
The $\sigma$'s are estimated from the Monte Carlo simulations
of the signal decays (see below).
As an example, we show in Fig.~\ref{fig:dedm}
the $\Delta E$ vs. $\Delta M$ distributions
of the candidate events for the decays $\ppio$ and $\p2pio$~\cite{ellipse}.

The numbers of events observed ($N_{ob}$) in the signal region
and the detection efficiencies ($\epsilon$)
are listed in Table~\ref{table:Results}.
The efficiencies are estimated from a Monte Carlo simulation.
In the Monte Carlo program, one $\tau$ lepton decays according to a
two- or three-body phase space distribution for the
mode of interest and the other $\tau$ lepton decays generically  
according to the KORALB $\tau$ event generator~\cite{KORALB}.
The detector response is simulated with the GEANT program~\cite{GEANT}.

The background ($N_{bg}$) is estimated from the sideband regions in the
$\Delta E$ vs. $\Delta M$ distribution assuming that the background shape is linear.
Each sideband is separated from the signal region by $6.0\sigma$
(see Fig.~\ref{fig:dedm} as an example).
The numbers of events observed are consistent with the background
expectations as shown in Table~\ref{table:Results}.
There is therefore no evidence for a signal.
To understand the origin of the background, we also estimate
the $\tau$ migration background using the KORALB program
and the hadronic background using the Lund program~\cite{Lund}.
The simulation programs can account for the background
and indicate that most of the background is from $\tau$ migration.
The large backgrounds in the decays $\ppio$ and $\p2pio$
originate from the copious decays $\tau^{-} \to \pi^{-} \pi^{0} \nu_{\tau}$
and $\tau^{-} \to \pi^{-} 2 \pi^{0} \nu_{\tau}$.

The upper limit on the branching fraction is related to the
upper limit $N$ on the number of signal events by
\begin{eqnarray}
\nonumber
{\cal B} =
\frac {N}{ 2 \epsilon N_{\tau\tau} {\cal B}_1 {\cal B}^m_{\pi^0}
{\cal B}^n_{\eta}}\ \ ,
\end{eqnarray}
where ${\cal B}_1$ is the inclusive 1-prong branching fraction~\cite{pdg},
${\cal B}_{\pi^0}$ (${\cal B}_{\eta}$) is the branching fraction~\cite{pdg} for
$\pi^0 \to \gamma\gamma$ ($\eta \to \gamma\gamma$), and $m$ ($n$)
is the number of $\pi^0$ ($\eta$) mesons in the final state.
The 90\% confidence level upper limits on the signal
are summarized in Table~\ref{table:Results}.
We estimate the upper limits using a Monte Carlo calculation,
which incorporates both the Poisson statistics of the signal and the
systematic errors.
The systematic errors include the statistical uncertainty in the
background estimate due to limited statistics in the sideband regions.
This statistical uncertainty is incorporated using Poisson
statistics~\cite{Poisson}.
All other sources of systematic errors are incorporated using 
Gaussian statistics. These include the uncertainties in the 
$\tau^+\tau^-$ cross section (1\%),
luminosity (1\%), track reconstruction efficiency (3\%),
photon detection efficiency (2.5\%), $p/\bar p$ detection efficiency (10\%),
branching fraction of $\eta\to\gamma\gamma$
(0.8\%)~\cite{pdg}, and the statistical uncertainties
in the detection efficiencies due to limited
Monte Carlo samples (1-2\% for the two-body
modes and 2-3\% for the three-body modes).
% The systematic error in the acceptance is estimated by assuming
% 100\% uncertainty in the 20\% lower detection efficiency for
% the modes involving an $\bar{p}$ due to
% the higher probability of $\bar{p}$ interacting in the material
% before the calorimeter, producing showers which cause
% the events to be rejected.
These uncertainties are added in quadrature in computing $N$.

In conclusion, we have searched for $\tau$ decays that
violate lepton and baryon numbers, but conserve baryon minus
lepton number. We find no evidence for a signal, resulting in
much improved upper limits for the two-body decays and first upper limits
for the three-body decays.

We gratefully acknowledge the effort of the CESR staff in
providing us with excellent luminosity and running conditions.
This work was supported by the National Science Foundation,
the U.S. Department of Energy, Research Corporation,
the Natural Sciences and Engineering Research Council of Canada,
the A.P. Sloan Foundation, the Swiss National Science Foundation,
and the Alexander von Humboldt Stiftung.

\begin{table}[htbp]
\begin{center}
\caption[]{Summary of detection efficiencies, signal yields, expected backgrounds, and
90\% C.L. upper limits on the signal yields and branching fractions.}
\vspace{0.1in}
\label{table:Results}
\begin{tabular}{lccccc}
Mode                    & $\pgamma$ & $\ppio$ & $\peta$ & $\p2pio$ & $\ppioeta$ \\
\hline
$\epsilon~(\%)$         & $\effa$   & $\effb$ & $\effc$ & $\effd$  & $\effe$    \\  
$N_{ob}$                & $\oba$    & $\obb$  & $\obc$  & $\obd$   & $\obe$     \\
$N_{bg}$                & $\bga$    & $\bgb$  & $\bgc$  & $\bgd$   & $\bge$     \\
$N$                     & 2.8       & 8.8     & 3.5     & 10.0     & 3.5        \\
$\BR~(10^{-6})$         & $\bra$    & $\brb$  & $\brc$  & $\brd$   & $\bre$ \\
\end{tabular}
\end{center}
\end{table}

\begin{figure}[p]
\centering
\centerline{\hbox{
\psfig{figure=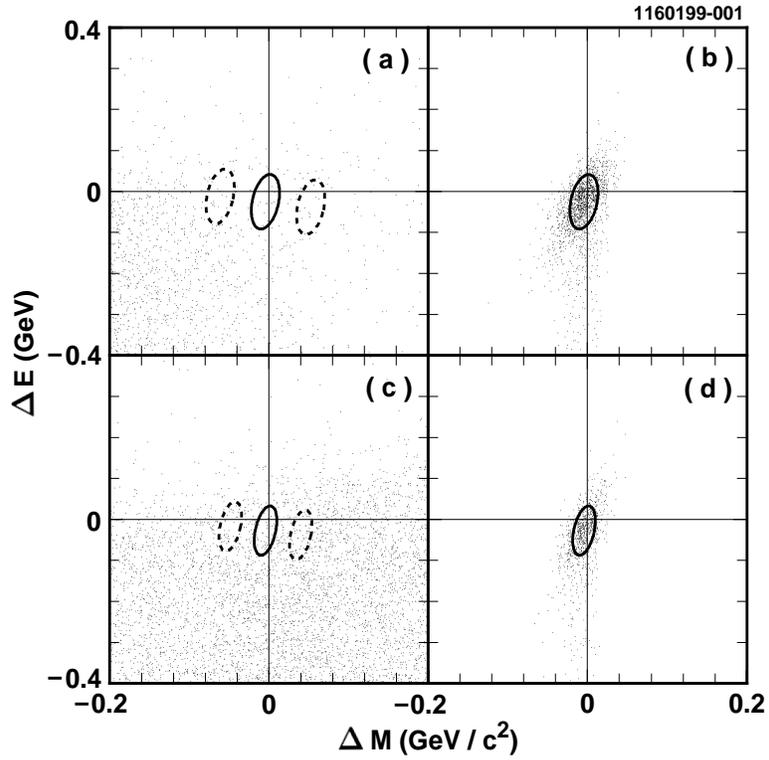,width=4.0in,height=4.0in}
}}
\vspace{0.25in}
\caption
{$\Delta$E vs. $\Delta$M distribution
of the data (a) and signal Monte Carlo (b) for the decay $\ppio$.
(c) and (d) show the corresponding distributions for $\p2pio$.
The normalization of the signal Monte Carlo is arbitrary.
The ellipses indicate the signal (solid) and sideband (dashed) regions.}
\label{fig:dedm}
\end{figure}

\begin{thebibliography}{99}
\bibitem{charge}
Throughout this paper, the charge conjugate state is implied.
\bibitem{ellipse}
The signal ellipse is slightly rotated and shifted from zero
in the $\Delta E$ vs. $\Delta M$ plane to account for
initial state radiation and shower leakage.
\end{thebibliography}
\end{document}